# High-throughput volumetric adaptive optical imaging using compressed time-reversal matrix

Hojun Lee[1,2,+], Seokchan Yoon[1,2,+], Pascal Loohuis[3], Jin Hee Hong[1], Sungsam Kang[1,2], and Wonshik Choi[1,2,*]

[1]*Center for Molecular Spectroscopy and Dynamics, Institute for Basic Science, Seoul 02841, Korea*

[2]*Department of Physics, Korea University, Seoul 02841, Korea*

[3]*Department of Applied Mathematics, University of Twente*

[+]*These authors contributed equally to this work.*

[*]*wonshik@korea.ac.kr*

## Abstract

Deep-tissue optical imaging suffers from the reduction of resolving power due to tissue-induced optical aberrations and multiple scattering noise. Reflection matrix approaches recording the maps of backscattered waves for all the possible orthogonal input channels have provided formidable solutions for removing severe aberrations and recovering the ideal diffraction-limited spatial resolution without relying on fluorescence labeling and guide stars. However, measuring the full input-output response of the tissue specimen is time-consuming, making the real-time image acquisition difficult. Here, we present the use of a time-reversal matrix, instead of the reflection matrix, for fast high-resolution volumetric imaging of a mouse brain. The time-reversal matrix reduces two-way problem to one-way problem, which effectively relieves the requirement for the coverage of input channels. Using a newly developed aberration correction algorithm designed for the time-reversal matrix, we demonstrated the correction of complex aberrations using as small as 2 % of the complete basis while maintaining the image reconstruction fidelity comparable to the fully sampled reflection matrix. Due to nearly 100-fold reduction in the matrix recording time, we could achieve real-time aberration-correction imaging for a field of view of 40×40 µm$^2$ (176×176 pixels) at a frame rate of 80 Hz. Furthermore, we demonstrated high-throughput volumetric adaptive optical imaging of a mouse brain by recording a volume of 128×128×125 µm$^3$ (568×568×125 voxels) in 3.58 s, correcting tissue aberrations at each and every 1 µm depth section, and visualizing myelinated axons with a lateral resolution of 0.45 µm and an axial resolution of 2 µm.

## Introduction

An arbitrary optical system interacting with light waves can be described by transmission and reflection matrices, as far as the linear light-matter interaction is concerned. The transmission (reflection) matrix of an optical system describes the transmitted (reflected) electric field (E-field) at all the possible detection channels for a set of orthogonal input channels. Due to the full characterization of the input-output response, the measured matrix can be considered a replica of a real optical system. Therefore, one can computationally process it as though a real experiment is being conducted. The knowledge of the matrix allows one to find ideal solutions in a variety of applications where lengthy experimental optimizations are required. Examples include focusing light[1,2], delivering images[3], and controlling transmission power[4] through scattering media based on the transmission matrix.

The reflection matrix, suitable for more realistic *in vivo* applications for which the detector cannot be placed on the transmission side, has provided exceptional opportunities for deep-tissue imaging[5]. The reflection matrix has also been exploited to focus light on a target embedded deep within strongly scattering media[6–8]. The wave correlation of the single-scattered waves in the reflection matrix was tailored to attenuate the effect of multiple light scattering[9]. A wavefront correction algorithm termed closed-loop accumulation of single scattering

(CLASS)[10] was developed based on the time-gated reflection matrix for separately identifying the input and output aberrations without the need for guide stars and in the presence of strong multiple scattering noise. This offers imaging deep within biological tissues with an ideal diffraction-limited resolution[11]. The singular value decomposition (SVD) was applied to a time-gated reflection matrix for retrieving a target image underneath strongly scattering media[12,13]. Recently, it has been demonstrated that the time-gated reflection matrix measured in the space domain made it possible to image a mouse brain through an intact skull inducing extreme aberrations[14]. Indeed, the reflection matrix approaches provide formidable solutions in the context of computational adaptive optics (AO) microscopy[15,16] in that they can deal with extremely severe aberrations with no need for fluorescence labeling and guide stars. In addition, this space-domain reflection matrix study proved that it can serve as a type of wavefront sensorless AO[17,18] that is combined with hardware correction of aberration by wavefront shaping devices such as a spatial light modulator and deformable mirror to realize ideal diffraction-limited multi-photon fluorescence imaging through an intact skull[14].

Despite these benefits, the matrix-based AO approach has been elusive in real-time bio-medical imaging applications. The recording of the full reflection matrix is a time-consuming process because the E-field map of the reflected wave must be measured for each illumination channel, as opposed to confocal imaging's requirement of point detection. Furthermore, the interferometric detection of the E-field is sensitive to the random phase drift, which can deteriorate the recorded reflection matrix in the dynamic samples. Sparse sampling of the matrix would be a potential solution, but this is accompanied by incomplete sampling of input channels. Considering that finding an object embedded within a scattering medium requires identifications of wave distortions in both the input and output pathways, insufficient sampling of the input channels can undermine the capability to resolve input distortions.

To overcome these issues, we consider a time-reversal matrix $RR^\dagger$ instead of the reflection matrix $R$. Here $R^\dagger$ represents the conjugate transpose of $R$. Unlike the reflection matrix itself, which describes the relationship between the input and output channels, the time-reversal matrix describes the phase-conjugated roundtrip process connecting the output channels to the output channels via the input channels. Essentially, this reduces the two-way problem with the reflection matrix to the one-way problem with the time-reversal matrix on condition that the input channels are orthogonal. There are two major benefits of dealing with the time-reversal matrix. It can maintain high fidelity in terms of retaining the information on the output channels even if the input channel coverage is much smaller than that of the complete set. Furthermore, it is not even necessary to know the basis of the input channels, making it robust to the random phase drift.

Here, we present a high-throughput volumetric AO imaging method termed a compressed time-reversal closed-loop accumulation of single scattering (CTR-CLASS), in which the previously developed CLASS algorithm was extended to a compressed time-reversal matrix constructed by a sparsely sampled reflection matrix for correcting the complex sample-induced aberrations with significantly reduced number of measurements. In this implementation, we took advantage of the time-reversal matrix and made use of dynamically varying unknown speckle patterns as input channels. We demonstrated that both the aberration map and object image can be retrieved using the number of speckle patterns as small as 2 % of the complete basis while maintaining comparable fidelity to that of the fully sampled matrix. Due to nearly 100-fold reduction of the matrix recording time, the CTR-CLASS has enabled real-time aberration-correction imaging for a field of view (FOV) of 40×40 µm$^2$ (176×176 pixels) at a frame rate of 80 Hz. We applied the developed method for the volumetric AO imaging of ex vivo mouse brain and visualized myelinated axons with a lateral resolution of 0.45 µm and axial resolution of 2 µm over a volume of 128×128×125 µm$^3$ (568×568×125 voxels) within a recording time of 3.58 s.

## Results

### Principle of CTR-CLASS

We consider the time-gated coherent imaging of a target object through a scattering medium in reflection geometry (Fig. 1**a**). For convenience, the optical layout is unfolded by flipping the reflection beam path over an object plane, making the layout analogous to transmission geometry. Since the scattering sample serves as a linear system with respect to the E-field in coherent imaging, the reflected wave can be described by a linear superposition of impulse response functions,

$$E_o(\mathbf{r}_o; \mathbf{r}_i) = \int P_o(\mathbf{r}_o; \mathbf{r}) O(\mathbf{r}) P_i(\mathbf{r}_i; \mathbf{r}) d^2 \mathbf{r}. \tag{1}$$

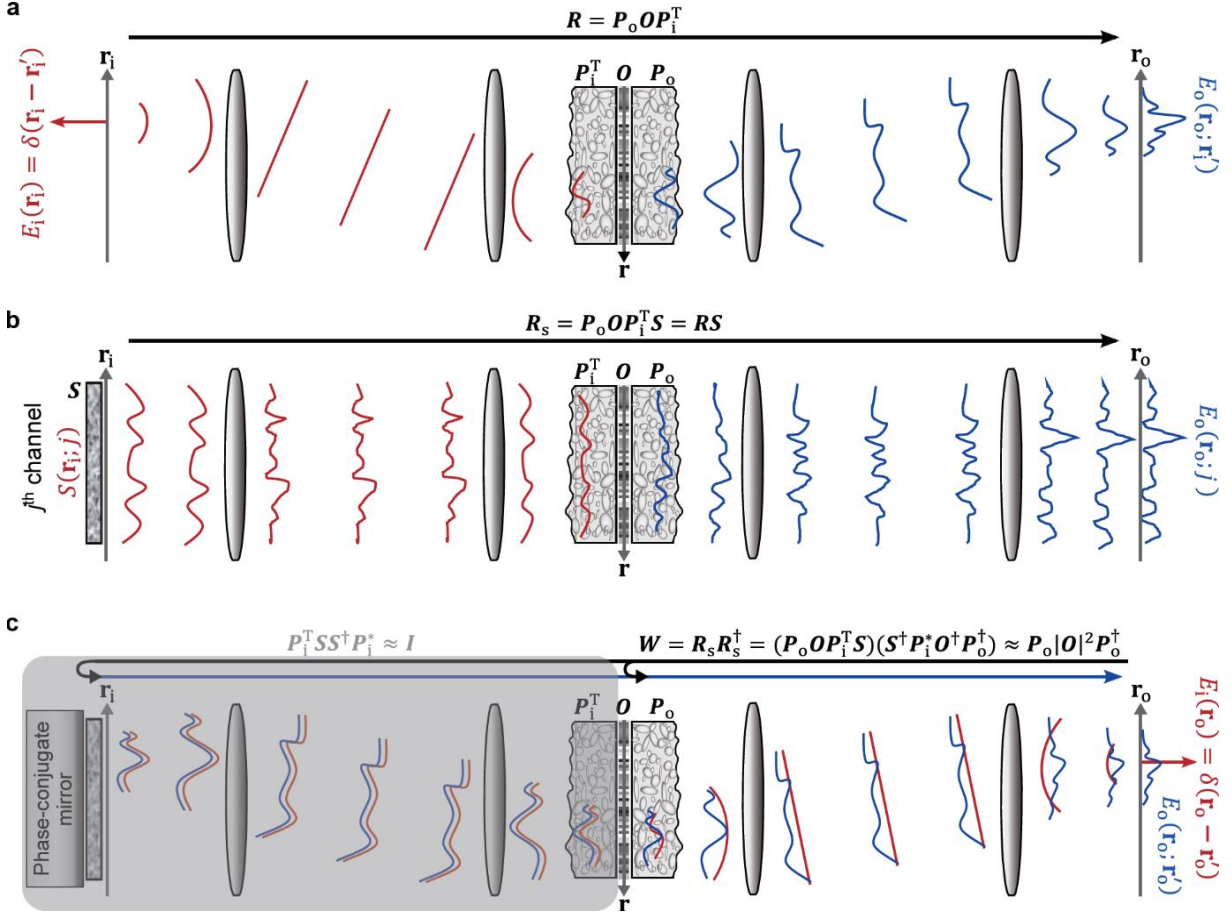

**Fig. 1 Schematics of imaging geometry and time-reversal process. a** Description of the reflection matrix $R$ in point illumination basis. $R[r_o; r'_i] = [E_o(r_o; r'_i)]$ is a set of E-field responses $E_o(r_o; r'_i)$ to impulse input fields $E_i(r_i) = \delta(r_i - r'_i)$. For clarity, the reflection pathway is unfolded to the transmission side. $O$ is the object's reflection coefficient matrix, and $P_{i(o)} = [P_{i(o)}(r; r_{i(o)})]$ is the transmission matrix of the scattering medium for wave propagation from position $r$ to $r_{i(o)}$. The red and blue curves represent incident and reflected waves, respectively. **b** Reflection matrix $R_S$ in the case of speckle illumination. $R_S[r_o; j] = [E_o(r_o; j)]$ is a set of E-field responses $E_o(r_o; j)$ for speckle input channels $S(r_i; j)$. $R_S$ can be written as $R_S = RS$, where $S[r_i; j] = [S(r_i; j)]$ is the input illumination matrix. **c** Geometric interpretation of the time reversal matrix. The time-reversal matrix $W[r_o; r'_o] = R_S R_S^\dagger$ can be interpreted as a reflection matrix describing a roundtrip process for a wave propagating from the output plane $r_o$ to the object plane $r$ and reflecting back to the output plane by the target object $|O|^2$.

Here, $E_o(\mathbf{r}_o; \mathbf{r}_i)$ is the time-gated E-field at position $\mathbf{r}_o$ on the output plane when a target object is illuminated by a point source located at position $\mathbf{r}_i$ on the input plane. $O(\mathbf{r})$ is the object function that represents complex reflection coefficients of the target object. Both the input and output planes are conjugate to the object plane whose spatial coordinate is $\mathbf{r}$. $P_{i(o)}(\mathbf{r}_{i(o)}; \mathbf{r})$ is the time-gated E-field point-spread-function (PSF) that maintains shift-invariance for the illumination (detection) path, i.e., $P_{i(o)}(\mathbf{r}_{i(o)}; \mathbf{r}) = P_{i(o)}(\mathbf{r}_{i(o)} - \mathbf{r})$. Respectively, they represent field distributions at the input (output) plane generated by a point source located at a position $\mathbf{r}$ on the object plane. Note that multiple-scattered waves also contribute to $E_o(\mathbf{r}_o; \mathbf{r}_i)$ in Eq. (1), but we excluded them for the sake of simplicity. As we reported earlier[10], the CLASS algorithm is developed in such a way to find object function even in the presence of multiple scattering noise. The same is true for our newly developed CTR-CLASS algorithm. In scattering matrix formalism, Eq. (1) can be represented by a time-gated reflection matrix $\boldsymbol{R}$ whose element is $E_o(\mathbf{r}_o; \mathbf{r}_i)$ for a column index $\mathbf{r}_i$ and row index $\mathbf{r}_o$. Based on Eq. (1), $\boldsymbol{R}$ can be decomposed as

$$\boldsymbol{R} = \boldsymbol{P}_o \boldsymbol{O} \boldsymbol{P}_i^T. \tag{2}$$

Here, $\boldsymbol{O}$ is a diagonal matrix whose diagonal element is $O(\mathbf{r})$. $\boldsymbol{P}_i$ and $\boldsymbol{P}_o$ are transmission matrices of the scattering medium covering the object plane whose matrix elements are $P_i(\mathbf{r}_i; \mathbf{r})$ and $P_o(\mathbf{r}_o; \mathbf{r})$, respectively. The superscript 'T' indicates the transpose of a matrix. For identical illumination and detection paths, input and output PSFs are the same due to the reciprocity principle in optics and thus satisfy the relation $P_i(\mathbf{r}'; \mathbf{r}) = P_o(\mathbf{r}'; \mathbf{r})$. However, this is not a necessary condition in the present study.

The matrix $\boldsymbol{R}$ in Eq. (2) can be directly measured by scanning the position $\mathbf{r}_i$ of the focused illumination and wide-field detection of the backscattered wave field $E_o(\mathbf{r}_o; \mathbf{r}_i)$ across $\mathbf{r}_o$. To obtain the full time-gated reflection matrix for a given FOV, it is necessary to scan a focused beam over the FOV with a lateral sampling interval of the diffraction-limited resolution, $\Delta x = \lambda/(2\text{NA})$, where $\lambda$ is the wavelength of light source, and NA is the objective numerical aperture. For a 2-dimensional (2D) FOV of size $L \times L$, the required number of sampling points for a complete sampling is given by $N = L/\Delta x$ which is the total number of orthogonal spatial modes for the given FOV, NA, and $\lambda$.

The reflection matrix can be measured by sending any complete set $\{E_i(\mathbf{r}_i; j)\}_{j=1}^N$ of $N$ illumination fields, instead of point-by-point scanning with a focused beam. One can measure the respective output field $E_o(\mathbf{r}_o; j)$ for each $j^{\text{th}}$ illumination and construct a reflection matrix $\boldsymbol{R}_m$ whose columns are assigned by the measured output E-fields $\{E_o(\mathbf{r}_o; j)\}_{j=1}^N$. In this case, the column and row indices of $\boldsymbol{R}_m$ are $j$ and $\mathbf{r}_o$, respectively. Then, the measured $\boldsymbol{R}_m$ is expressed as

$$\boldsymbol{R}_m = \boldsymbol{R} \boldsymbol{E}_i, \tag{3}$$

where $\boldsymbol{E}_i$ is an illumination matrix constructed by $\{E_i(\mathbf{r}_i; j)\}_{j=1}^N$ in the same way as $\boldsymbol{R}_m$. The reflection matrix of the sample can be obtained by multiplying the measured matrix $\boldsymbol{R}_m$ by the inverse of $\boldsymbol{E}_i$, i.e., $\boldsymbol{R} = \boldsymbol{R}_m \boldsymbol{E}_i^{-1}$. This requires knowledge of the illumination fields.

Here, we propose the use of a set of $M$ unknown random speckle illumination patterns, $\{S(\mathbf{r}_i; j)\}_{j=1}^M$ as a special case of illumination basis. In particular, we consider the case in which $M$ is significantly smaller than $N$. The time-gated E-field image $E_o(\mathbf{r}_o; j)$ is recorded for each $j^{\text{th}}$ speckle illumination, and the sparsely-sampled reflection matrix $\boldsymbol{R}_s$ is then constructed using $\{E_o(\mathbf{r}_o; j)\}_{j=1}^M$ as a matrix element (Fig. 1**b**). Therefore, $\boldsymbol{R}_s$ becomes an $N$-by-$M$ matrix with column and row indices of $j$ and $\mathbf{r}_o$, respectively. The matrix $\boldsymbol{R}_s$ can be expressed as

$$\boldsymbol{R}_s = \boldsymbol{P}_o \boldsymbol{O} \boldsymbol{P}_i^T \boldsymbol{S}, \tag{4}$$

where $\boldsymbol{S}$ is an $N$-by-$M$ illumination matrix constructed by $\{S(\mathbf{r}_i; j)\}_{j=1}^M$. To realize aberration correction and image reconstruction without a prior knowledge of the illumination pattern, we consider a CTR matrix, $\boldsymbol{W} = \boldsymbol{R}_s \boldsymbol{R}_s^\dagger$. By inserting Eq. (4) in $\boldsymbol{W}$, the matrix is expressed as

$$\boldsymbol{W} = (\boldsymbol{P}_o \boldsymbol{O} \boldsymbol{P}_i^T \boldsymbol{S})(\boldsymbol{S}^\dagger \boldsymbol{P}_i^* \boldsymbol{O}^\dagger \boldsymbol{P}_o^\dagger) \approx \boldsymbol{P}_o \boldsymbol{O}_I \boldsymbol{P}_o^\dagger, \tag{5}$$

where the superscript '*' indicates the complex conjugate. Here, $\boldsymbol{P}_i^T \boldsymbol{S} \boldsymbol{S}^\dagger \boldsymbol{P}_i^*$ is approximated as an identity matrix $\boldsymbol{I}$. This approximation is valid when the speckle illumination patterns are not correlated with each other, i.e., $\sum_{j=1}^M S(\mathbf{r}_i; j) S^*(\mathbf{r}_i'; j) \approx \delta_{\mathbf{r}_i, \mathbf{r}_i'}$, and thus $\boldsymbol{S} \boldsymbol{S}^\dagger \approx \boldsymbol{I}$. In addition, it is necessary to assume that $\boldsymbol{P}_i^T \boldsymbol{P}_i^* = \boldsymbol{P}_i \boldsymbol{P}_i^\dagger \approx \boldsymbol{I}$. This approximation is valid when $\boldsymbol{P}_i$ is either shift-invariant or sufficiently complex. Note that $\boldsymbol{S}$ and $\boldsymbol{P}_{i(o)}$ are properly normalized such that $\langle \sum_j |S(\mathbf{r}_i; j)|^2 \rangle_{\mathbf{r}_i} = 1$, $\langle \sum_{\mathbf{r}_i} |S(\mathbf{r}_i; j)|^2 \rangle_j = N$, and $\langle \sum_{\mathbf{r}_i} |P_i(\mathbf{r}_i; \mathbf{r})|^2 \rangle_{\mathbf{r}} = 1$, where the bracket notation denotes an average over the variable in the subscript, for the sake of simplicity. As will be discussed below, the assumption of $\boldsymbol{P}_i^T \boldsymbol{S} \boldsymbol{S}^\dagger \boldsymbol{P}_i^* \approx \boldsymbol{I}$ is generally valid for sufficiently large $M$. Then, $\boldsymbol{W}$ is reduced to $P_o O_I P_o^\dagger$, where $O_I$ denotes $|O|^2$, a diagonal matrix with its diagonal element given by the reflectance of the object, $|O(\mathbf{r})|^2$.

There are a few major benefits of considering $W$ instead of $R$. At first, the matrix $S$ describing the illumination patterns is removed in $W$, thereby eliminating the need to know the illumination speckle patterns. This makes it possible to send an arbitrary choice of illuminations, such as dynamic speckle patterns generated by a rotating diffuser, and thus it is no longer necessary to scan pre-defined positions of point illumination using scanning mirrors. Furthermore, $W$ is greatly simplified such that only $P_o$ and $O_I$ remain to be identified. Imperfection in illumination and detection optics often causes discrepancy between $P_i$ and $P_o$ in the reflection geometry. $P_i$ and $P_o$ are intrinsically different in the transmission geometry. However, in $W$, it is not necessary to consider $P_i$. Another critical benefit is the possibility of sparse sampling. The condition $SS^\dagger \approx I$ satisfies even when $M$ is extremely small. In contrast, if there is significant downsampling in the focused illumination, both the ability to identify the aberration and the imaging fidelity are significantly degraded.

Physical interpretation of the time-reversal matrix $W$ is given in Fig. 1c. $W = R_s R_s^\dagger$ is a successive operation of $R_s^\dagger$ and $R_s$. By the $R_s^\dagger$ operation, a spherical wave (red curves) emitted from a point source at a position $\mathbf{r}_o'$ on the output plane propagates in the backward direction through the object ($P_i^* O^\dagger P_o^\dagger$) followed by a fictious scattering layer whose transmission matrix is $S^\dagger$. Afterwards, the $R_s$ is applied such that the reflected wave (blue curves) returns to the scattering layer ($S$) and the object ($P_o O P_i^T$) in the forward direction to arrive at the output plane. Here, the important point is that the operation indicated by the shaded gray box ($P_i^T SS^\dagger P_i^*$) serves as a phase-conjugation mirror when $SS^\dagger \approx I$, i.e., the illumination speckles are sufficiently orthogonal. In other words, a point source emanating from an object plane comes back to the same position via its travel through $P_i^T SS^\dagger P_i^*$.

This eliminates the need to consider the input aberration $P_i$ and illumination patterns $S$. As a result, the matrix $W = R_s R_s^\dagger$ can be interpreted as a time-gated reflection matrix describing an imaging system that images a reflective object with the reflectance $O_I$ through a scattering medium with an input transmission matrix of $P_o^*$ and an output transmission matrix of $P_o$. The whole process becomes an $N$-by-$N$ square matrix with its column and row indices both corresponding to $\mathbf{r}_o$.

The concept of the time-reversal matrix was initially introduced for selective focusing in acoustics[19,20] and then has been intensively studied in microwaves[21,22] and optics[7,12,23]. In these previous studies, successive operation of the time-reversal matrix $W = RR^\dagger$ was used to find an eigenvector of $R$ with the largest eigenvalue. This operation induces the selective wave to focusing on the object with the highest reflectance. In contrast, we considered the CTR matrix $W = R_s R_s^\dagger$ and introduced the matrix decomposition $W = P_o O_I P_o^\dagger$ to find the unknown object function ($O_I$) embedded within a scattering medium inducing optical aberrations ($P_o$).

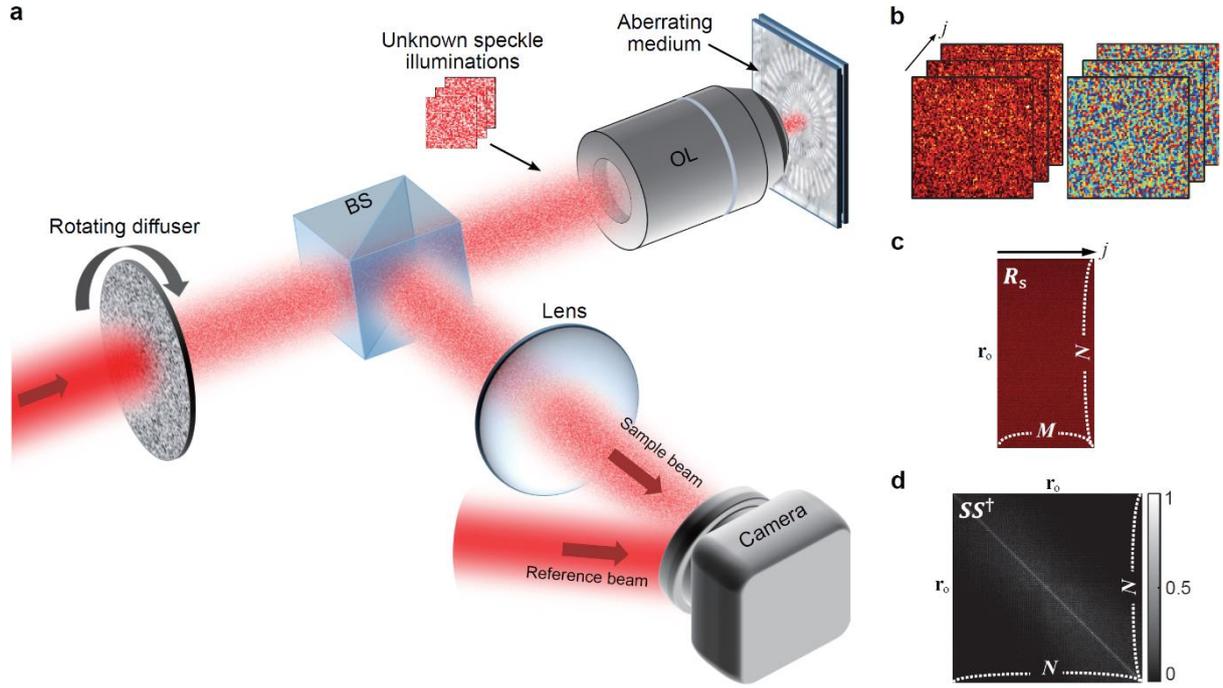

**Fig. 2 Experimental setup for recording a compressed time-reversal matrix. a** Schematic of the experimental setup. BS: beamsplitter, OL: objective lens. A Ti:Sapphire pulsed laser is used as a low-coherence light source. Random speckle patterns produced by a rotating diffuser illuminate a sample. The off-axis holograms between the sample wave and a plane reference wave are recorded by a camera. Optics for reference beam path is omitted for brevity. **b** Representative E-field amplitude (left) and phase (right) images of the sample wave. The size of each image is 40×40 μm$^2$, composed of $N$ = 7,744 orthogonal modes. **c** Sparsely-sampled timed-gated reflection matrix $\mathbf{R}_S$ reconstructed by a set of $M$ = 700 E-field images. **d** The matrix $\mathbf{SS}^\dagger$ obtained from an experimentally measured $\mathbf{S}$ by placing a mirror at the sample plane

### Experimental setup of CTR-CLASS microscopy

The schematic of the experimental setup is shown in Fig. 2**a** for recording a CTR matrix $\mathbf{W}$. The basic configuration is a low-coherence wide-field interferometric microscope, but a sample is illuminated by random speckle fields while the reference wave is a planar wave. A custom-built wavelength-tunable Ti:Sapphire pulsed laser (center wavelength of 800 - 900 nm, bandwidth of 30 nm) was used as a low-coherence light source. An optical diffuser mounted on a motorized rotation stage was inserted at a conjugate image plane in the illumination path to produce uncorrelated random speckle fields for the sample wave. Backscattered sample waves from the target were captured by an objective lens (Nikon, x60, NA 1.0) and delivered to a high-speed CMOS camera (Photron, FASTCAM mini UX100) placed at a conjugate image plane. A reference plane wave was introduced at the camera to generate the off-axis low-coherence interferogram, from which we obtained the time-gated E-field of the backscattered sample wave (see Supplementary Information Note I for the detailed setup). Figure 2**b** shows three representative E-field amplitude and phase images of the sample under different speckled illuminations. To obtain a complete $N$-by-$N$ reflection matrix for a FOV having $N$ orthogonal modes, $N$ E-field images must be acquired for an orthogonal set of $N$ illumination fields, where each image has a total of $N$ orthogonal pixels. However, we recorded only $M$ ($<N$) E-field images using random speckle illuminations to reduce the acquisition time and reconstruct a sparsely sampled $N$-by-$M$ reflection matrix $\mathbf{R}_S$ as shown in Fig. 2**c**.

For efficient sparse sampling of the reflection matrix, it is important to minimize the correlation between the speckle fields as much as possible. For a given camera exposure time and frame rate, the angular velocity of the rotating diffuser was carefully selected to minimize the spatial correlation between two consecutive speckle patterns (Supplementary Information Note II). To justify the validity of $\mathbf{SS}^\dagger \approx \mathbf{I}$, we experimentally measured the $\mathbf{S}$ matrix by placing a mirror at the sample plane. Figure 3**d** shows $\mathbf{SS}^\dagger$ matrix obtained using $M$ = 700 speckled illuminations for a FOV of 40×40 μm² with a diffraction-limited resolution of 450 nm, resulting in the total number of orthogonal modes, $N = 88 \times 88 = 7,744$. The $SS^\dagger$ was nearly diagonal matrix with the ratio between off-diagonal and diagonal elements of $\lesssim 0.1$.

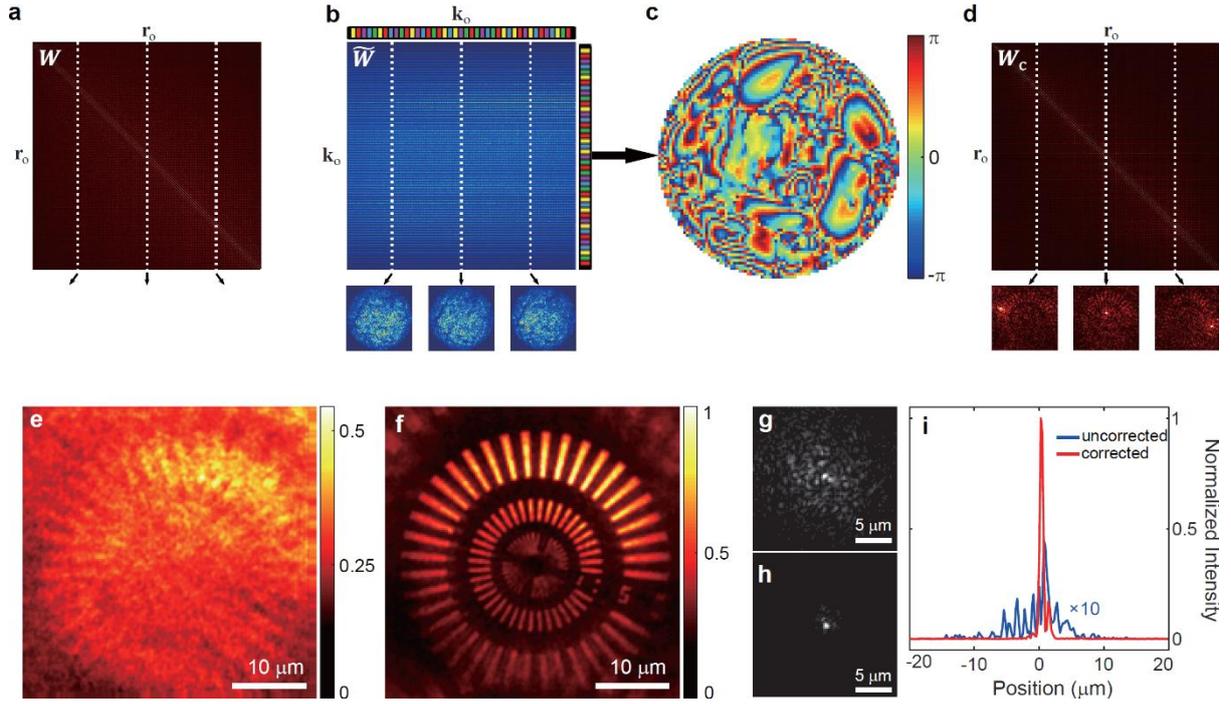

**Fig. 3 Wavefront aberration correction using time-reversal matrix. a** Fully-sampled time-reversal matrix $W = R_s R_s^\dagger$ ($CR = 1$) for a Siemens star resolution target underneath a highly aberrating medium. **b** Time-reversal matrix $\widetilde{W}$ in k-space, obtained by the Fourier transform of $W$. **c** Wavefront aberration map $\phi_o(k_o)$ at the output pupil plane, identified by the CLASS algorithm. Color bar indicates the phase in radians. **d** Aberration-corrected time-reversal matrix $W_c$. The subset images in (**a, b,** and **d**) are 2D E-field images corresponding to columns marked by dashed lines. **e, f** Uncorrected and corrected images of the Siemens star reconstructed from $\widetilde{W}$ and $\widetilde{W}_c$, respectively. Images are normalized by the maximum intensity in the corrected image. **g, h** Uncorrected and corrected intensity-PSFs, respectively. **i** Line profiles of the uncorrected and corrected intensity-PSFs in (**g, h**).

### Aberration correction with CTR-CLASS

To demonstrate the high-throughput data acquisition and aberration correction capabilities of the CTR-CLASS microscopy, we imaged a homemade Siemens star target covered by a 600-μm-thick plastic layer introducing strong optical aberrations. The laser operated at a center wavelength of 900 nm, and had a coherence length of ~12 μm. For a FOV of 40×40 μm² ($N = 88 \times 88$ pixels), $M$ speckled E-field images of the target were imaged by the high-speed camera operating at a frame rate of 12,500 Hz with an exposure time of 20 μs.

We define the compression ratio that indicates the degree of sparse sampling as $CR = M/N$. Figure 3**a** shows the time-gated $W$ matrix constructed by speckle patterns for $CR = 1$. In the absence of aberration ($P_o = I$), the matrix $W \approx P_o O_I P_o^\dagger$ is almost diagonal because it is reduced to $O_I$. In the presence of aberration and scattering, the signal in the diagonal spreads out to the off-diagonal elements. For space-invariant aberrations, $P_o$ becomes a Toeplitz matrix. To identify the space-invariant (i.e., angle-dependent) aberration and reconstruct an aberration-corrected image, the basis of $W$ is changed to the spatial-frequency domain by taking the Fourier transform for both the column and row bases. The resulting matrix denoted as $\widetilde{W}$ is shown in Fig. 3**b**. $\widetilde{W}$ is expressed as

$$\widetilde{W} = \widetilde{P}_o \widetilde{O}_I \widetilde{P}_o^\dagger, \tag{6}$$

where $\widetilde{O}_I$ and $\widetilde{P}_o$ are the Fourier transformations of $O_I$ and $P_o$, respectively. The matrix $\widetilde{P}_o(\mathbf{k}_o; \mathbf{k})$ represents the transfer matrix between the object and output planes. When the output aberration is space-invariant, i.e.,

$P_o(\mathbf{r}_o; \mathbf{r}) = P_o(\mathbf{r}_o - \mathbf{r})$, $\widetilde{\boldsymbol{P}}_o$ becomes a diagonal matrix. Its diagonal element is given by $\tilde{P}_o(\mathbf{k}_o; \mathbf{k}_o) = e^{i\phi_o(\mathbf{k}_o)}$, where $\phi_o(\mathbf{k}_o)$ is the output pupil phase map.

This change of basis allows us to apply the CLASS algorithm[10] on the time-reversal matrix $\widetilde{\boldsymbol{W}}$, which identifies an aberration correction matrix $\widetilde{\boldsymbol{P}}_c$ that maximizes the total intensity of the reconstructed image. This was done by applying $\widetilde{\boldsymbol{P}}_c$ to $\widetilde{\boldsymbol{W}}$ to obtain the corrected time-reversal matrix, $\widetilde{\boldsymbol{W}}_c = \widetilde{\boldsymbol{P}}_c \widetilde{\boldsymbol{W}} \widetilde{\boldsymbol{P}}_c^*$. $\widetilde{\boldsymbol{P}}_c$ is a diagonal matrix whose elements are given by $\tilde{P}_c(\mathbf{k}_o; \mathbf{k}_o) = e^{i\phi_c(\mathbf{k}_o)}$. The CLASS algorithm iteratively finds $\tilde{P}_c(\mathbf{k}_o)$ in such a way that the sum of the diagonal elements of $\boldsymbol{W}_c$ is maximized. Here, $\boldsymbol{W}_c$ is the inverse Fourier transform matrix of $\widetilde{\boldsymbol{W}}_c$. At the $n^{\text{th}}$ iteration, the $n^{\text{th}}$ correction pupil function $\tilde{P}_c^{(n)}(\mathbf{k}) = e^{i\phi_c^{(n)}(\mathbf{k})}$, target spectrum $\tilde{O}_I^{(n)}(\Delta\mathbf{k})$, and time-reversal matrix $\widetilde{\boldsymbol{W}}^{(n)}$ are calculated as

$$\phi_c^{(n)}(\mathbf{k}_o) = \arg\{\frac{1}{N}\sum_{\Delta\mathbf{k}\neq 0} \widetilde{W}^{(n-1)}[\mathbf{k}'_o + \Delta\mathbf{k}; \mathbf{k}'_o] \cdot \tilde{O}_I^{(n-1)*}(\Delta\mathbf{k})\} \tag{7}$$

$$\widetilde{W}^{(n)}[\mathbf{k}_o; \mathbf{k}'_o] = \tilde{P}_c^{(n)}(\mathbf{k}_o)\widetilde{W}^{(n-1)}[\mathbf{k}_o; \mathbf{k}'_o]\tilde{P}_c^{(n)*}(\mathbf{k}'_o), \tag{8}$$

$$\tilde{O}_I^{(n)}(\Delta\mathbf{k}) = \frac{1}{N}\sum_{\mathbf{k}'_o} \widetilde{W}^{(n)}[\mathbf{k}'_o + \Delta\mathbf{k}; \mathbf{k}'_o], \tag{9}$$

where $\Delta\mathbf{k} = \mathbf{k}_o - \mathbf{k}'_o$. Note that the target spectrum $\tilde{O}_I^{(n)}(\Delta\mathbf{k})$ is reconstructed by synthesizing all the spatial frequency spectra covered by illumination angles $\mathbf{k}'_o$, resulting in a cut-off frequency $|\Delta\mathbf{k}/(2\pi)| = 2NA/\lambda$. The iteration starts with the initial conditions of $\tilde{P}_c^{(0)}(\mathbf{k}_o) = 1$ and $\widetilde{W}^{(0)} = \widetilde{W}$ and continues until the root-mean-square (RMS) error of $\phi_c^{(n)}$ at $n = n_{\max}$ becomes less than a predefined value. The final output correction phase function is found by accumulating all the preceding correction phases, $\phi_c(\mathbf{k}_o) = \sum_{n=1}^{n_{\max}} \phi_c^{(n)}(\mathbf{k}_o)$. The identified wavefront aberration $\phi_c(\mathbf{k}_o)$ is shown in Fig. 3**c**, and the corrected CTR matrix $\boldsymbol{W}_c$ is shown in Fig. 3**d**. The intensity images before and after the aberration correction were reconstructed from $\widetilde{\boldsymbol{W}}$ and $\widetilde{\boldsymbol{W}}_c$, respectively. The uncorrected image shown in Fig. 3**e** is blurry and hardly recognizable, while object structures are sharply resolved in the corrected image in Fig. 3**f**. Both the imaging resolution and the signal-to-background ratio (SBR) in Fig. 3**f** were significantly improved compared to those in the uncorrected image. We quantified the degree of aberration correction by measuring the normalized intensity profiles of PSFs before and after the aberration correction (Figs. 3**g-i**). From the line profiles of the PSFs, the increase in the Strehl ratio, which is the ratio between the peak intensities before and after the correction, was measured to be 10. The measured full-width at half-maximum (FWHM) of the aberration-corrected PSF was about 450 nm, which is the diffraction-limit spatial resolution of the system.

We evaluated the performance of image recovery depending on *CR* to determine the minimum achievable *CR*. Reconstructed images and aberration phase maps for various *CR* are shown in Fig. 4**a**. The first row shows the reconstructed intensity images normalized by *M* for *CR* = 0.5, 0.1, 0.02, and 0.017, and the second row shows the corresponding aberration phase maps. The identified aberration maps were almost identical all the way into the high spatial frequency range, although *CR* was significantly reduced. Diffraction-limited high-resolution images could be successfully restored for $CR \geq 0.02$ ($M \geq 155$). Considering the camera frame rate of 12,500 Hz, it took only 12.4 ms to record the CTR matrix $W$ for $CR = 0.02$, setting the highest achievable aberration-correction image frame rate to 80 Hz for a FOV of 40×40 μm² (88×88 pixels). If the $CR$ was further reduced to less than 0.02, the image reconstruction failed to find the correct aberration map and object image. The line profiles along the white dotted lines on the reconstructed images in Fig. 4**a** are compared in Fig. 4**b** to quantify the image quality. Interestingly, neither the image contrast nor the spatial resolution of the reconstructed image was diminished by the reduction of $CR$, so far as the image reconstruction was successful. This means that even the severe decrease in $CR$ does not hamper the performance of aberration correction. We compared the residual root-mean-square (RMS) wavefront errors of the identified wavefront maps relative to the aberration map obtained from the full reflection matrix (blue dots in Fig. 4**c**). The residual RMS wavefront errors remained nearly constant regardless of $CR$. This indicates that aberrations were properly corrected even with the small $M$.

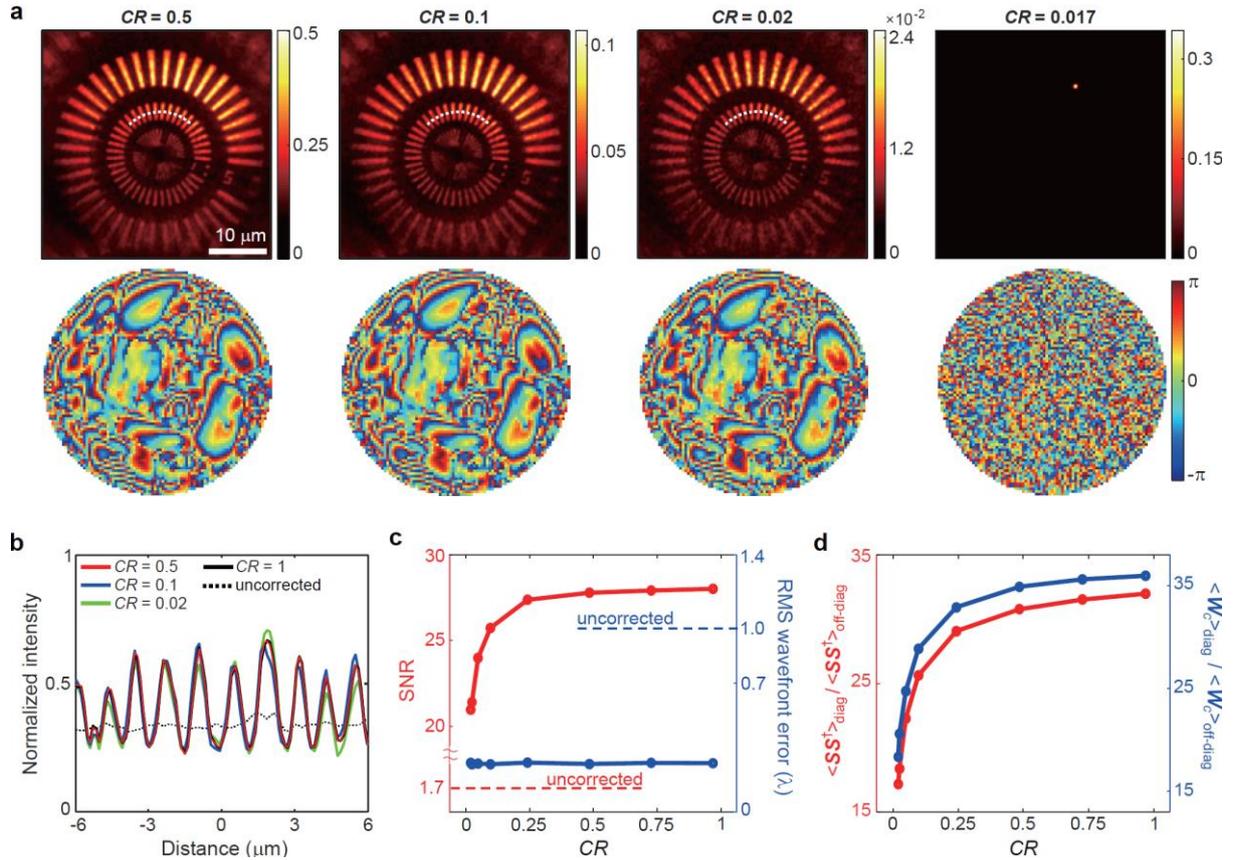

**Fig. 4 Image reconstruction fidelity depending on the compression ratio ($CR$). a** Aberration-corrected images (upper row) and corresponding aberration maps (lower rows) for various $CR$ values. **b** Line profiles along the white dashed lines in (**a**). **c** SNR (red dots) and residual RMS wavefront error (blue dots) depending on $CR$. **d** Ratio between the averaged diagonal elements and off-diagonal elements of $SS^\dagger$ (red dots) and $W_c W_c^\dagger$ (blue dots) depending on $CR$. The solid curves in (**c, d**) are linear interpolations of the data.

An important figure of merit in imaging is the signal-to-noise ratio (SNR), which is defined by the ratio of the mean intensity of the target image and the standard deviation of random background noise. We plotted the SNR as a function of $CR$ in Fig. 4**c** (red dots) and observed that the SNR decreased with $CR$. The SNR fits well with $\sqrt{CR}$ because the signal grows with $N_s$, and the background fluctuations grow with $\sqrt{M}$ after the aberration correction.

The SNR is determined by the ratio of the diagonal to off-diagonal elements of the $SS^\dagger$ matrix (red dots in Fig. 4**d**). With the decrease of $CR$, this factor was reduced such that the condition $P_i^T SS^\dagger P_i^* \simeq I$ is gradually violated. The off-diagonal elements of $SS^\dagger$ are random complex numbers with a mean amplitude of $1/\sqrt{M}$ and a random phase factor of $e^{i\varphi_{\text{rand}}}$. This is confirmed by the fact that the diagonal to off-diagonal ratio of the $SS^\dagger$ matrix shown in Fig. 4**d** fits well with $\sqrt{M}$. Then, the term $\boldsymbol{P}_i^T \boldsymbol{SS}^\dagger \boldsymbol{P}_i^*$ can be statistically expressed as $\boldsymbol{I} + \frac{1}{\sqrt{M}}\boldsymbol{N}$, where $\boldsymbol{N}$ is a random matrix, whose elements are given by $a_{\text{rand}} e^{i\varphi_{\text{rand}}}$ with a mean amplitude of $a_{\text{rand}}$ unity. Considering the non-zero random off-diagonal elements of $\boldsymbol{SS}^\dagger$, the time-reversal matrix $\boldsymbol{W}$ in Eq. (6) is written as

$$\boldsymbol{W} = \boldsymbol{P}_o \boldsymbol{O}_I \boldsymbol{P}_o^\dagger + \boldsymbol{W}_N, \tag{10}$$

where $\boldsymbol{W}_N = \frac{1}{\sqrt{M}} \boldsymbol{O} \boldsymbol{N} \boldsymbol{O}^\dagger$ is the noise term introduced by $\boldsymbol{N}$. The matrix element of $\boldsymbol{W}_N$ is a random phasor with a standard deviation given by $\sigma_I = \tilde{O}_I(0)/\sqrt{M}$, where $\tilde{O}_I(0) = \langle O_I(\mathbf{r})\rangle_\mathbf{r}$ is a DC spectral component given by the average reflectance of the target. This inherent noise term originates from the illumination inhomogeneities due to the insufficient number of speckle patterns. It acts analogous to random noise due to multiple scattering in a turbid medium. If the CTR-CLASS algorithm finds the aberration properly, the aberration-corrected CTR matrix $\boldsymbol{W}_c$ is given by $\boldsymbol{W}_c = \boldsymbol{P}_c \boldsymbol{P}_o \boldsymbol{O}_I \boldsymbol{P}_o^\dagger \boldsymbol{P}_c^\dagger + \boldsymbol{P}_c \boldsymbol{W}_N \boldsymbol{P}_c^\dagger$. The signal after aberration correction is then given by $\alpha_S \langle O_I(\mathbf{r})\rangle_\mathbf{r}$, where $\alpha_S$ is the Strehl ratio of the residual PSF after the aberration correction is obtained by the diagonal elements of $\boldsymbol{P}_c \boldsymbol{P}_o$. The standard deviation of the noise due to $\boldsymbol{P}_c \boldsymbol{W}_N \boldsymbol{P}_c^\dagger$ is almost identical to that of $\boldsymbol{W}_N$, as the elements of $\boldsymbol{W}_N$ are random phasors. Therefore, the SNR of the reconstructed image is given by SNR = $\alpha_S \sqrt{M}$. The signal ratio between the diagonal and off-diagonal elements of the $\boldsymbol{W}_c$ matrix (blue dots in Fig. 4**d**) is the signal to background ratio, which in turn is equal to SNR. We found that this ratio fits well with $\sqrt{M}$, confirming the validity of our analysis.

Image reconstruction was successful up to a *CR* value of 0.02, but it was shown to fail with a further decrease of *CR*. As explained above, the reduction in *CR* introduces random noise. When the CTR-CLASS algorithm calculates cross-correlation between angular spectrum images to estimate the aberration phase map, and the noise induced by the sparse sampling adds phase error. Below a certain threshold *CR* value, the noise level becomes too high for the algorithm to retrieve the correct aberration phase map. Essentially, the minimum achievable *CR* value is determined by the SNR, i.e. the ratio between the signal from the object and sparsity-induced noise, Therefore, appropriate choice of *CR* value is necessary depending on the reflectance of the sample for optimal image acquisition speed.

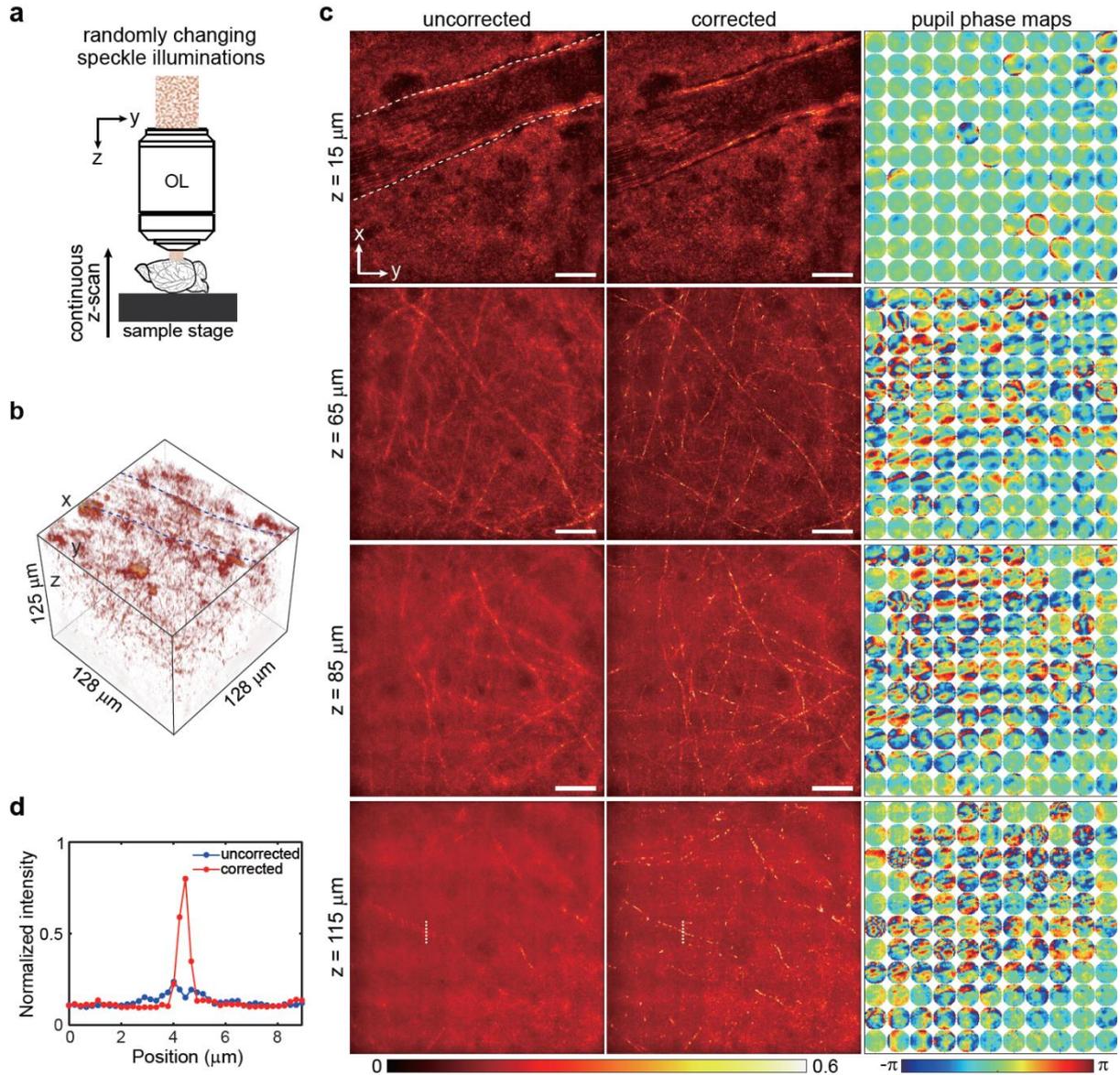

**Fig. 5 Aberration-free volumetric high-speed imaging of a mouse brain. a** Imaging configuration. The mouse brain tissue was continuously scanned along the z-axis at a constant speed while dynamically varying speckle patterns illuminated the specimen. **b** Reconstructed 3D volumetric image of brain tissue with transverse field of view of 128 × 128 µm and depth range of 125 µm, composed of 568 × 568 ×125 voxels. The two white dashed curves indicate wall boundaries of a blood vessel with a diameter of ~30 µm, located close to the brain surface. **c** Left-hand and middle columns show MIP images with and without aberration correction at various depths, respectively. MIP range: 6 µm for z = 15, 65, and 85 µm, and 20 µm for z = 115 um. The right-hand column shows reconstructed output pupil phase maps for 11×11 subregions. The white dashed curves in the uncorrected image at z = 15 µm indicate wall boundaries of a blood vessel near the brain surface. The radius of each circle corresponds to a numerical aperture of NA = 0.94. The images at each depth were normalized by the maximum intensity in the corrected image. Scale bar: 20 µm. **d** Line profiles along the vertical white dotted lines in the images at z = 115 µm in (**b**).

## Volumetric aberration-correction imaging of a mouse brain

With the CTR-CLASS microscope, we demonstrated the high-throughput aberration-corrected volumetric imaging of myelinated axons in an *ex vivo* mouse brain. In this demonstration, the pulsed laser operated at a center wavelength of 848 nm with a coherence length of 8 µm. Typically, a set of E-field images constituting a reflection matrix are recorded for each fixed depth to maintain the input and output planes throughout the measurements.

This depth-wise matrix recording slows down the volumetric imaging. Here, we employed continuous depth scanning to speed up the volume coverage. Since the output planes are continuously varying, we added numerical propagation steps to synchronize the output planes prior to the application of CTR-CLASS algorithm. The synchronization of the input planes is not necessary as there is no need to know the used input channels. To experimentally implement this concept, a whole mouse brain was mounted on a motorized stage and continuously scanned along the z-axis at a constant speed of 35 µm/s while dynamically varying speckle patterns illuminated the specimen (Fig. 5a). The E-field images of the backscattered waves were recorded by the high-speed camera at a frame rate of 5,000 Hz. Therefore, there was 7 nm depth difference between the neighboring frames. The camera exposure time (50 µs) and the angular velocity of the rotating diffuser (210 degree/s) were carefully selected to ensure that the spatial correlation of the speckle patterns between the consecutive frames was less than 0.1 (See Supplementary Information Note II). For the total image acquisition time of 3.58 s, we obtained a series of 17,891 E-field images with a frame size of 128×128 µm$^2$ (300×300 pixels), spanning a depth range of 125 µm from the surface of the brain. The angular spectra of the obtained E-field images are filtered by applying a binary pupil mask with NA = 0.94. Then, we finally obtained E-field images with a size of 128×128 µm$^2$ (284×284 pixels) and diffraction-limited resolution of 0.45 nm,

To reconstruct a 3D volume image, we first prepared depth-corrected E-field images at 125 depths spaced 1 µm apart. To retrieve an aberration-corrected image at each given depth z, E-field images taken within a range of z ± 4 µm numbering 1147 were numerically propagated to the depth z by adding appropriate quadratic spectral phase factors in their angular spectra. To deal with position-dependent aberrations, the depth-corrected E-field images were divided into 11×11 subregions, and the CTR matrix for each subregion was separately constructed using these images. Finally, we retrieved aberration-corrected 2D images by applying the CLASS algorithm to the CTR matrices at individual subregions in all depth z. The reconstructed 3D volume image over 128×128×125 µm$^3$ (568×568×125 voxels) is shown in Fig. 5b (See Supplementary Video 1). Note that the depth-dependent defocus due to refractive index mismatch between immersion water and the tissue causes the separation of the objective focus and coherence volume resulting in blurred images[24]. We could find and compensate the depth-dependent defocus by numerically propagating the E-field images such that the total intensity of reconstructed images without aberration correction was maximized. The depth-dependent defocus was about 3 µm at z = 100 µm, which was less than the coherence length (8 µm) of light source. Representative section images at various depths are shown in Fig. 5c. The left-hand column shows maximum intensity projection (MIP) of the reconstructed images without aberration correction, whereas the middle column shows corresponding MIP images after aberration correction. The right-hand column shows identified output pupil phase maps for 11×11 subregions. There were no significant aberrations at z = 15 µm except for spherical aberration. Up to a depth of 50 µm, the aberration-corrected images were almost identical to those without correction. As the imaging depth was increased, the aberration maps became more complex, and myelinated fibers in the uncorrected images became blurred due to the inhomogeneity within the tissue. Specifically, there was a blood vessel with a diameter of ~30 µm located close to the surface of the brain. The white dashed curves in the uncorrected image at z = 15 µm in Fig. 5c indicate wall boundaries of the blood vessel. We observed that the blood vessel induced pronounced aberrations such that the aberration maps in areas under the vessel were more complex than those in other areas. In addition, correlation between aberration maps of neighboring subregions decreased rapidly, suggesting that the isoplanatic patch size was reduced with depth. At the depth of 115 µm, myelinated axons were almost invisible without aberration correction. Intensity line profiles along the white dotted lines in the images at z = 115 µm are shown in Fig. 5d. Comparing the line profiles, we observed that the CTR-CLASS can recover a nearly diffraction-limit resolution of ~0.45 µm (the minimum thickness of myelinated fiber in FWHM) and high-contrast images of myelinated fibers (up to ~7-fold increase in signal-to-background ratio). The axial resolution measured from cross-sections of myelin fibers along z-axis was ~2 µm.

## Discussion

The reflection matrix containing full optical input-output response of a scattering medium has offered robust image reconstruction in comparison with conventional adaptive optics approaches relying on partial information. In particular, it enables the correction of extremely complex aberrations in stringent conditions where there are strong multiple scattering noise and no guide stars available. As a trade-off, the matrix recording is too time-consuming to perform real-time imaging. Throughout our study, we demonstrated that the use of a time-reversal matrix, instead of the reflection matrix, can be a solution for the high-throughput volumetric imaging equipped with all the benefits of the reflection matrix approaches. We proved that the time-reversal matrix approach can maintain the fidelity of aberration correction and image reconstruction using as small as 2 % of the full basis sampling. Due to nearly 100-fold reduction of the matrix recording time, we could achieve aberration-correction imaging for a 2D FOV of 40×40 µm$^2$ at a frame rate of 80 Hz. Furthermore, we realized the volumetric imaging of a mouse brain over a volume of 128×128×125 µm$^3$ in 3.58 s with a lateral resolution of 0.45 µm and an axial resolution of 2 µm throughout all the voxels including the areas underneath a blood vessel.

The proposed method presents a noteworthy conceptual advance. It is a new discovery that the time-reversal matrix can be highly compressed in terms of illumination channel coverage as long as the covered channels are orthogonal. We found that it is not even necessary to know what the illumination channels were. These conceptual findings naturally led to the advances in practicality. In addition to the reduction of illumination channel coverage, there is no need to perform time-consuming pre-calibration to gain prior knowledge on illumination field. It is no longer necessary to concern the phase stability among the E-field images. This enabled us to use dynamically varying random speckle patterns for illumination, instead of laser beam scanning by carefully aligned scanning mirrors, which greatly simplifies the experimental setup. We also presented novel volumetric image processing algorithm that replaced previous depth-wise angular scanning with continuous depth scanning in conjunction with dynamic speckle illuminations. We introduced the depth-correction step where the E-field images taken at various depths were propagated to the target depth. This increases the number of images to be used for constructing a time-reversal matrix, which effectively increases the volumetric imaging speed. All these benefits of using the compressed time-reversal matrix come with a price to pay. The sparsity of the illumination channel coverage introduces random noise equivalent to multiple scattering noise. Therefore, achievable imaging depth is reduced relative to the full sampling by the amount of sparsity-induced noise.

High-throughput volumetric imaging equipped with aberration correction capability for every depth section allows detailed mapping of microstructures deep within tissues. This will lead to accurate quantification of structural and molecular information in various biological systems. Therefore, the presented method will find its use for a wide range of studies in life science and medicine including the myelin-associated physiology in neuroscience, retinal pathology in ophthalmology and endoscopic disease diagnosis in internal organs. Due to the high-speed measurement of tissue aberration, it can also serve as wavefront sensing AO to provide aberration information for the hardware aberration correction. This will help to improve the imaging depth of fluorescence and nonlinear imaging modalities such as multi-photon microscopy, super-resolution microscopy, and coherent Raman microscopy.

## Materials and methods

### Animal Preparation

Adult (over 8 weeks) C57BL/6 mice were deeply anesthetized with an intraperitoneal injection of ketamine/xylazine (100/10 mg/kg) and decapitated. After the scalp and skull were removed, the brain was fixed with 4 % paraformaldehyde at 4°C overnight and washed with phosphate buffered saline (PBS) three times. For imaging, the fixed brain was stuck to a plastic dish and immersed in PBS. All animal experiments were approved by the Korea University Institutional Animal Care & Use Committee (KUIACUC-2019-0024).

## References

1. Popoff, S. M. *et al.* Measuring the Transmission Matrix in Optics: An Approach to the Study and Control of Light Propagation in Disordered Media. *Phys. Rev. Lett.* **104**, 100601 (2010).

2. Vellekoop, I. M. & Mosk, A. P. Focusing coherent light through opaque strongly scattering media. *Opt. Lett.* **32**, 2309–2311 (2007).

3.  Popoff, S., Lerosey, G., Fink, M., Boccara, A. C. & Gigan, S. Image transmission through an opaque material. *Nat. Commun.* **1**, 81 (2010).

4.  Kim, M. *et al.* Maximal energy transport through disordered media with the implementation of transmission eigenchannels. *Nat. Photonics* **6**, 581–585 (2012).

5.  Yoon, S. *et al.* Deep optical imaging within complex scattering media. *Nat. Rev. Phys.* **2**, 141–158 (2020).

6.  Choi, Y. *et al.* Measurement of the Time-Resolved Reflection Matrix for Enhancing Light Energy Delivery into a Scattering Medium. *Phys. Rev. Lett.* **111**, 243901 (2013).

7.  Popoff, S. M. *et al.* Exploiting the Time-Reversal Operator for Adaptive Optics, Selective Focusing, and Scattering Pattern Analysis. *Phys. Rev. Lett.* **107**, 263901 (2011).

8.  Jeong, S. *et al.* Focusing of light energy inside a scattering medium by controlling the time-gated multiple light scattering. *Nat. Photonics* **12**, 277–283 (2018).

9.  Kang, S. *et al.* Imaging deep within a scattering medium using collective accumulation of single-scattered waves. *Nat. Photonics* **9**, 253–258 (2015).

10. Kang, S. *et al.* High-resolution adaptive optical imaging within thick scattering media using closed-loop accumulation of single scattering. *Nat. Commun.* **8**, 2157 (2017).

11. Kim, M. *et al.* Label-free neuroimaging in vivo using synchronous angular scanning microscopy with single-scattering accumulation algorithm. *Nat. Commun.* **10**, 3152 (2019).

12. Badon, A. *et al.* Smart optical coherence tomography for ultra-deep imaging through highly scattering media. *Sci. Adv.* **2**, e1600370 (2016).

13. Badon, A. *et al.* Distortion matrix concept for deep optical imaging in scattering media. *Sci. Adv.* **6**, eaay7170 (2020).

14. Yoon, S., Lee, H., Hong, J. H., Lim, Y.-S. & Choi, W. Laser scanning reflection-matrix microscopy for aberration-free imaging through intact mouse skull. *Nat. Commun.* **11**, 5721 (2020).

15. Adie, S. G., Graf, B. W., Ahmad, A., Carney, P. S. & Boppart, S. A. Computational adaptive optics for broadband optical interferometric tomography of biological tissue. *Proc. Natl. Acad. Sci.* **109**, 7175 LP – 7180 (2012).

16. Liu, Y.-Z. *et al.* Computed optical interferometric tomography for high-speed volumetric cellular imaging. *Biomed. Opt. Express* **5**, 2988–3000 (2014).

17. Booth, M. J., Neil, M. A. A., Juškaitis, R. & Wilson, T. Adaptive aberration correction in a confocal microscope. *Proc. Natl. Acad. Sci.* **99**, 5788 LP – 5792 (2002).

18. Débarre, D. *et al.* Image-based adaptive optics for two-photon microscopy. *Opt. Lett.* **34**, 2495–2497 (2009).

19. Prada, C. & Fink, M. Eigenmodes of the time reversal operator: A solution to selective focusing in multiple-target media. *Wave Motion* **20**, 151–163 (1994).

20. Prada, C., Manneville, S., Spoliansky, D. & Fink, M. Decomposition of the time reversal operator: Detection and selective focusing on two scatterers. *J. Acoust. Soc. Am.* **99**, 2067–2076 (1996).

21. Tortel, H., Micolau, G. & Saillard, M. Decomposition of the Time Reversal Operator for Electromagnetic Scattering. *J. Electromagn. Waves Appl.* **13**, 687–719 (1999).

22. Micolau, G., Saillard, M. & Borderies, P. DORT method as applied to ultrawideband signals for detection of buried objects. *IEEE Trans. Geosci. Remote Sens.* **41**, 1813–1820 (2003).

23. Kim, D.-Y., Jeong, S., Jang, M., Lee, Y.-R. & Choi, W. Time-gated iterative phase conjugation for efficient light energy delivery in scattering media. *Opt. Express* **28**, 7382–7391 (2020).

24. Ben Arous, J. *et al.* Single myelin fiber imaging in living rodents without labeling by deep optical coherence microscopy. *J. Biomed. Opt.* **16**, 116012 (2011).

**Acknowledgements**

This work was supported by the Institute for Basic Science (IBS-R023-D1).

**Author contributions**

S.Y. conceived the idea. W.C. supervised the project. H.L. and S.Y. performed the experiments. P.L., H.L., and S.Y. wrote the Matlab code and analyzed the experimental data. J.H.H. prepared biological samples. W.C., S.Y., H.L., and S.K. wrote the manuscript. All authors discussed the results and commented on the paper.